\documentclass[a4paper,aps,prl,showpacs,amssymb,twocolumn,superscriptaddress]{revtex4}
\usepackage{graphicx,t1enc}

\begin{document}
\date{\today}

\title{The Social Behavior and the Evolution of Sexually Transmitted Diseases}

\author{Sebastián Gonçalves}
\email{sgonc@if.ufrgs.br} \affiliation{Instituto de Física
Universidade Federal do Rio Grande do Sul, Caixa Postal 15051,
90501-970 Porto Alegre RS, Brazil}
\author{Marcelo Kuperman}
\email{kuperman@cab.cnea.gov.ar} \affiliation{Centro Atómico
Bariloche and Instituto Balseiro, 8400 S. C. de Bariloche,
Argentina}

\begin{abstract}
{\normalsize We introduce a model for the evolution of sexually
transmitted diseases, in which the social behavior is incorporated
as a determinant factor for the further propagation of the
infection. The system may be regarded as a society of
agents where in principle anyone can sexually
interact with any other one in the population.
Different social behaviors are reflected in a distribution of sexual
attitudes ranging from the more conservative to the more
promiscuous. This is measured by what we call the promiscuity
parameter. In terms of this parameter, we find a critical
behavior for the evolution of the disease. There is a threshold
below what the epidemic does not occur. We relate this
critical value of the promiscuity to what epidemiologist call
the basic reproductive number, connecting it with the other
parameters of the model, namely the infectivity and the infective
period in a quantitative way.
We consider the possibility of subjects be grouped in couples.
In this contribution only the homosexual case is analyzed.}
\end{abstract}

\maketitle

\section{Introduction}
In the recent years, many mathematical models of  social phenomena
have  been formulated in order to describe a wide variety of
phenomena~\cite{Weid}. A particular interest was shown by the
evolution of epidemic processes when the structure of the
underlying society is taken into account~\cite{KA,MN,SV,ZC}. In
many  of these models the existence of a threshold value that
determines the further evolution of a nucleus of infection was
verified~\cite{SV2,Newm}. On the contrary,  it is possible to find
other situations where the absence of threshold was
shown~\cite{SV2}. Nevertheless, the interesting feature in all of
these models is that they incorporate a social aspect not taken
into account in previous epidemiological analysis. In  the
present work, we analyze a particular family of diseases, those
sexually transmitted such as AIDS, Hepatitis B, Syphilis, etc.  We
will focus on a particular aspect of the society very related to
the propagation of these diseases: the sexual behavior. We will
consider a society composed by sexually active individuals and
with different behavioral patterns ranging from the most
conservative or stable one, those who only have sexual
intercourse with their stable and unique mate, to the most
promiscuous who are continuously willing to change their sexual partner.
The dynamic of the disease will be associated to the $SIR$ case,
where, at a given time, each individual in the population can be
in one of the following three stages: susceptible ($S$), infective ($I$),
and refractory or removed ($R$). A susceptible
individual can become infective through contagion by an infective
individual. Once an element has been infective, it enters a cycle
that, after a fixed infection time, ends when the element
reaches the refractory (or removed in case of death, as in the present
model) state. At this stage the individual cannot be infected again nor infect
the others, therefore the
infection ultimately leads to definitive removal of elements from
the susceptible population. This family of models has been used
to describe the dynamics of well-known infectious diseases, such
as AIDS, rabies, and black death, by means of mean field
approximation~\cite{Murr}.

\section{The model}
Our model of the society conforms to a set of $N$ sexually active
subjects, which in principle are grouped in couples with
the exception of a controlled proportion of singles ($\rho_{sng}$).
Initially a ``promiscuity'' value  $p_i$ is randomly assigned
to each agent from a semi-Gaussian distribution of width $\hat p$.
$p_i$ determines the individual's tendency to
dismiss his stable mate and go out --or just to go out in the case
of the singles-- to look for an occasional intercourse.
In terms of the model the individual promiscuity $p_i$ is the
probability of trying to meet somebody else on each opportunity,
determined by the time step of the simulation.
Then, a susceptible individual can become infective, after
a sexual intercourse with an infective one, with probability $\beta$.
The infective individuals remain infectious for a period $\tau$,
after what they are removed by death.

Every time step each individual must choose to have sex with his
couple or with somebody else, the latter occurs with probability
$p_i$.
Those who have decided to go out, must chose a partner at random
and if the latter has decided to go out too, the occasional couple
is made. Therefore there is no social structure {\it a priori}.
The web of contacts is constructed dynamically during the
simulation depending on the percentage of singles and $\hat p$.
Thus, a random proportion of subjects, from those who have decided to go out, are
reorganized in temporary couples every time step. Those who were
abandoned by their couples and decided no to look for another
partner will not have sex during this time step. The same is true
for the singles who decided not to go out. Eventually, an
individual who goes out to look for somebody else may not have
success remaining sexually inactive.
In the present model, once an individual
is removed, no new members are added to the society.
This more complex situation will be analyzed in a further work.

Summarizing, the model has four parameters:
$\rho_{sng}$, $\hat p$, $\tau$, and $\beta$
The first one controls the proportion of singles,
the second one is the width of the semi-Gaussian distribution of $p_i$,
while the latter's two are the same for all individuals.
As long as we assume that all agents are
--except of the promiscuity and the marital status-- identical
(i.e. the infectivity between any
two subjects is symmetrical), the model may be regarded as
representing the spread of HIV/AIDS in a male homosexual community
with no distinction between passive and active partners.

\section{Numerical results}
As was stated in the Introduction our results are for fixed
number ($N$) of subjects, all sexually active --in
practice an individual could be sexually low active if
single and with low $p_i$.
No new members are incorporated during
the dynamics and no other cause of death apart from the sexually
transmitted disease is considered. Therefore the number of
susceptibles ($S$), infectives ($I$), and removeds ($R$) is conserved: $N
= S + I + R$. Most of the simulation are performed for 100000
subjects population with only one individual initially infective.
A hundred realizations (different runs starting with a different
shuffling of the $p_i$ among individuals) were considered to make
statistics.
In order to give the simulations a human dimension one time step
represents a day, and we make runs up to 5yr-100yr of simulation
time, depending on the selected parameters (namely $\tau$).

\begin{figure}[ht]
\includegraphics[width=8.cm, clip=true]{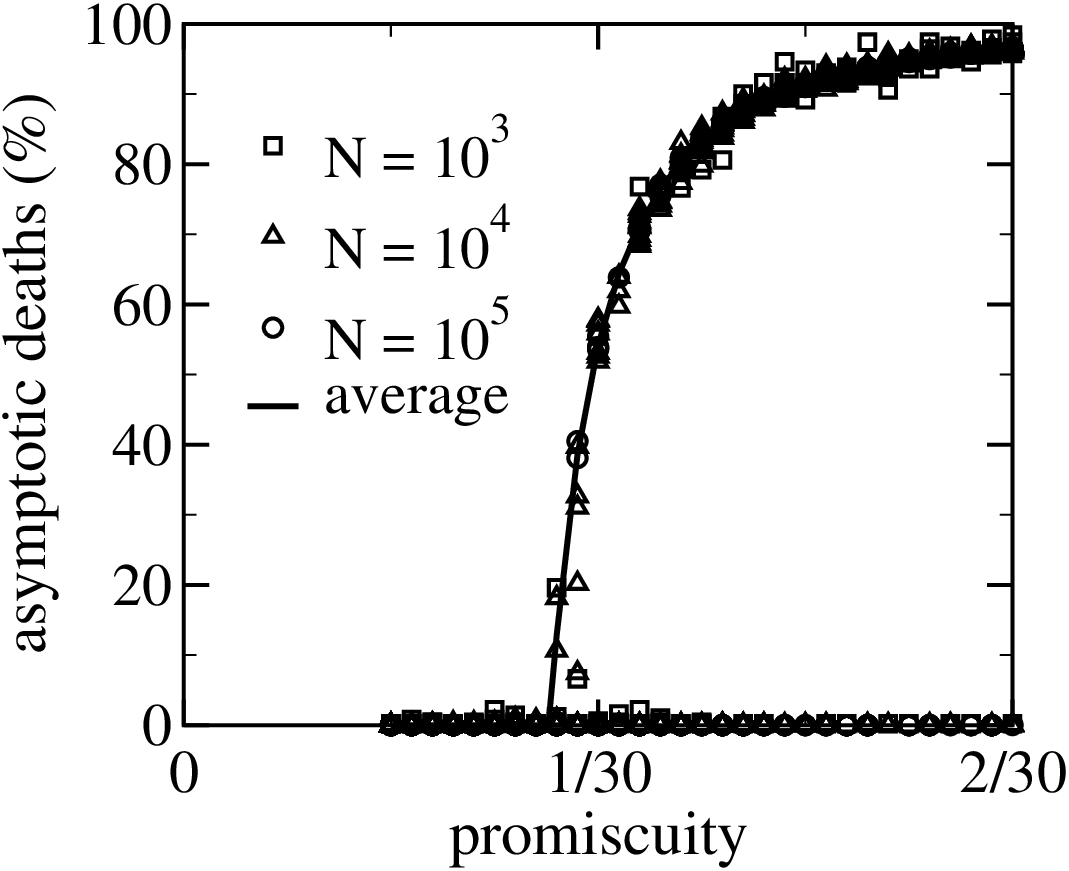}
\caption{
Asymptotic percentage of death as a function of the promiscuity
distribution width $\hat p$. Parameters are $\tau=1yr$ and $\beta=1$.
Notice the onset of the epidemic at the critical value
$\hat p_c = 0.9/30$. Numerical results obtained with $N=$~1000, 10000, and 100000 subjects,
100 realization and for $\rho_{sng}=0$ case (all subjects paired).}
\label{hiv}
\end{figure}

If we take for example $\rho_{sng}=0$ (all subjects paired),
$\hat p=1/30$, $\tau=1~yr$, and $\beta=1~day^{-1}$,
the resulting time
evolution of the infectives, $I(t)$, and susceptibles, $S(t)$,
is similar to what can be obtained solving the classical
$SIR$ differential equations~\cite{Murr}.
However, the similarities are only qualitative, because
in our scheme different runs for the same parameters
give different positions for the infectives maximum peak occurrence
for example. Even more, a small percentage of the realizations
does not produce an epidemic; i.e the infection dies out taking
just few individuals. This is an unexpected result from the point
of view of mean field standard continuous models, however it does
not seem unreasonable at all, considering that people, behavior
and randomness are involved in real life. Thus statistically there
is a low but non zero probability for the infection not to evolve
with these parameters, while the standard SIR model would give an
epidemic as unique, deterministic outcome. Statistically, these
cases should be considered as fluctuations.

The results for several runs with the above parameters,
but for $\hat p$ varying in the range $1/60-1/15$,
and for different populations sizes are presented in
Fig.~\ref{hiv}. There it can be seen that the probability
for an epidemic to occur, changes abruptly from zero
to a finite value, at the critical promiscuity $\hat p_c = 0.9/30$.
We can see too that there is a non-zero probability for non-epidemic at values
of $\hat p > \hat p_c$ as denoted by the points at the horizontal axis,
showing an asymptotic very low number of death.
At the same time, the asymptotic number of death increases very fast
with  $\hat p > \hat p_c$, for realizations that produce epidemics.
The line in Fig.~\ref{hiv} is the average of asymptotic deaths
calculated over all realizations which gave an epidemic
(\% of asymptotic death > 1\%). In what follows we show this average
instead of plotting the individuals runs.
Probably, the most interesting feature found in this work
-not previously reported- is the epidemic threshold in terms of the
promiscuity.  The critical value of the population promiscuity, $\hat p_c$,
depends obviously on the values of the other parameters
($\tau$, $\beta$, and $p_{sng}$) as we will see.

In the same Fig.~\ref{hiv} it can be noted that the obtained results are
not an effect of the finite size of the sample. Several
realizations on different sized population are plotted
there, and the coincidence of all data is quite evident.

Another aspect to take into account is the possibility of the
epidemic evolution of being critically affected by the number of
initially infected individuals, $I(0)$. Though it is clear that
for a huge proportion of initially infectives, the evolution
will drastically change, allowing for a small increment in the
number of them, namely from 1 to 5, in samples of
10000 subjects --and even in a small population of 1000 subjects--
gives the same results.

\begin{figure}[ht]
\includegraphics[width=8.cm, clip=true]{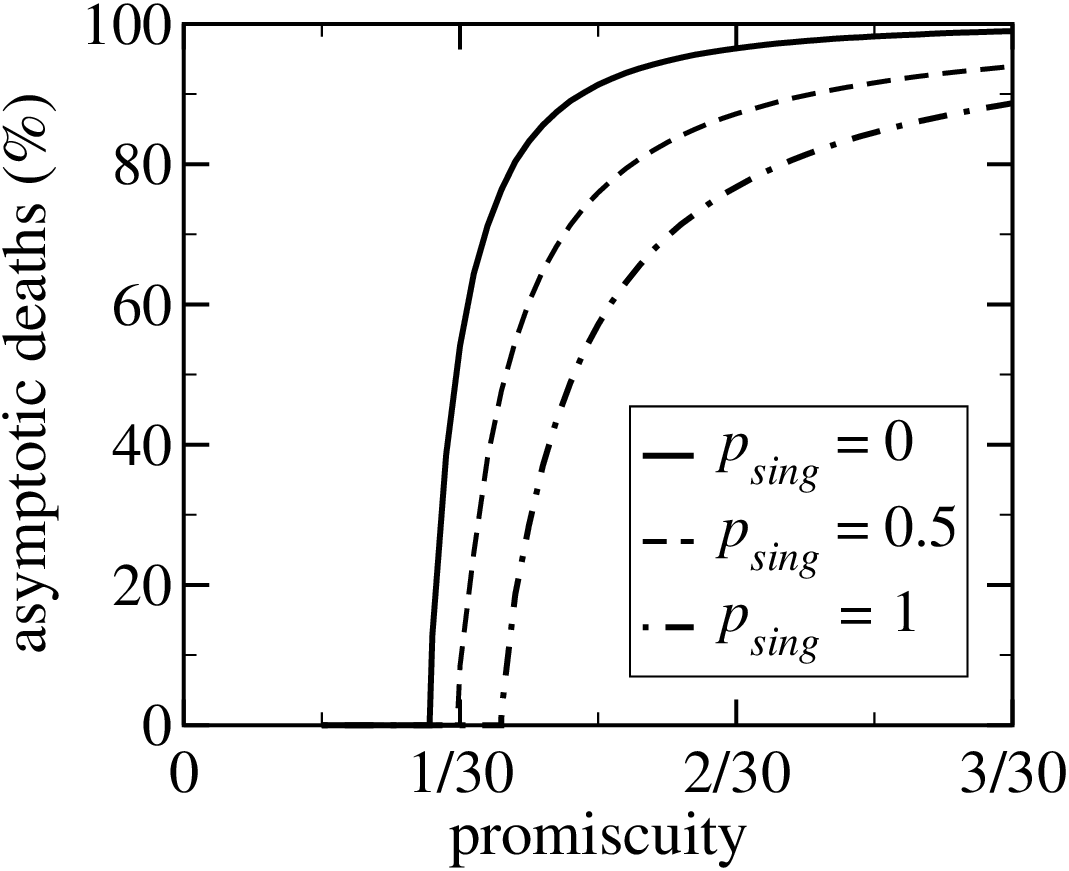}
\caption{
Asymptotic percentage of death as a function of distribution
promiscuity width for three different values of $\rho_{sng}$
($\beta=1~day^{-1}$, $\tau=1~yr$).
Numerical results for 100000 subjects and 100 realizations.}
\label{sing}
\end{figure}

Up to here the results presented are for $\rho_{sng}=0$,  i.e.
the non singles case. How the results depend on this late
parameter? Performing simulations for three different cases,
i.e. $\rho_{sng}=0$, $0.5$, $1$, we can see that the qualitative
behavior of the results is the same, with a displacement to higher
values of $\hat p_c$ as $\rho_{sng}$ increases. The results are plotted in
Fig.~\ref{sing}, where the shift in the value of $\hat p_c$ is observed,
and also a slower evolution to high values of death as $\hat p$ increases.
This is an expected result: once one of the members of the
couple gets infective, sooner or later the other gets too;
therefore, the more people are in couples, the more effective
is the propagation of the epidemic. We have to remark here
that in the present model the individual promiscuity is
independent of the marital status.

\begin{figure}[ht]
\includegraphics[width=8.cm, clip=true]{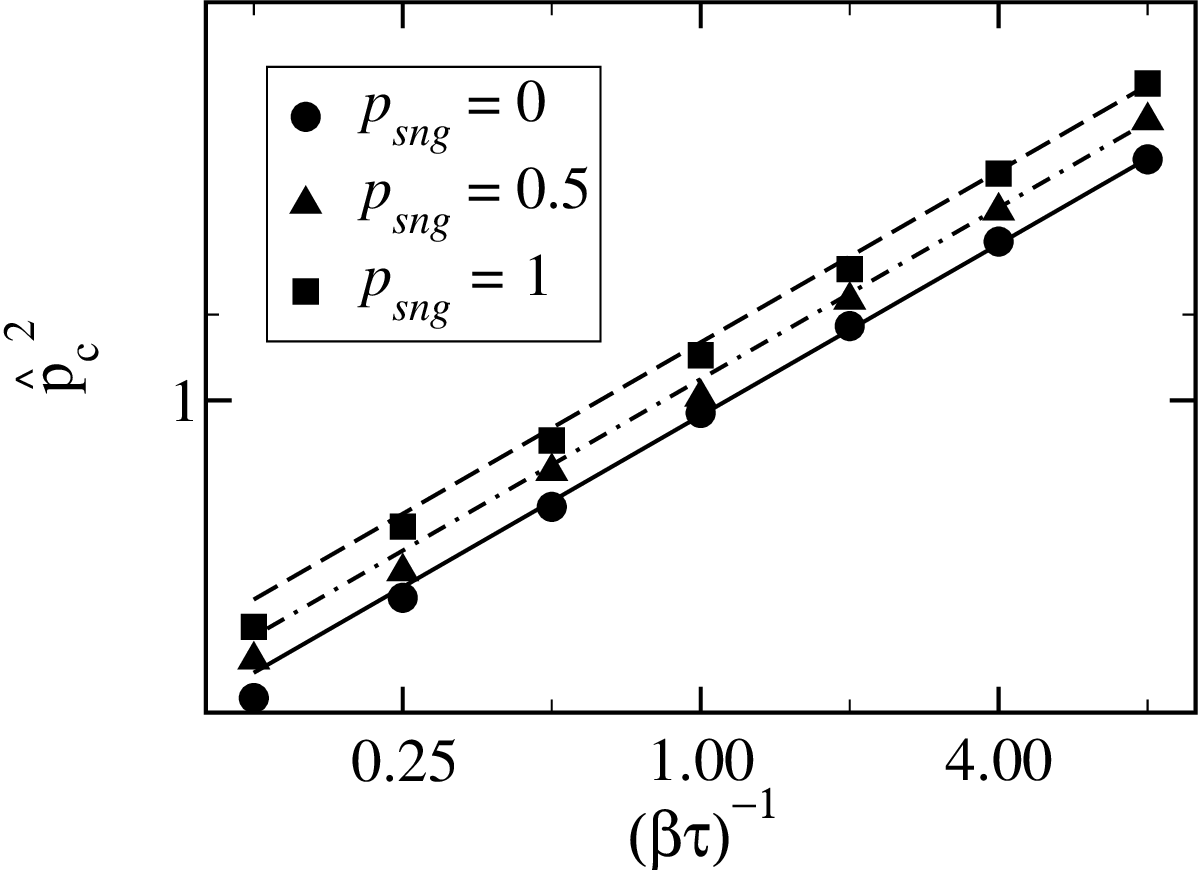}
\caption{
Squared critical value of the distribution promiscuity width
($\hat p^2_c$) as a function of the inverse product of $\beta$
and $\tau$, for three different values of $\rho_{sng}$.
Promiscuity is a per month probability, where $\beta$
is a per contact probability; $\tau$ is in years.}
\label{pc}
\end{figure}

It is known from classical $SIR$ models (see~\cite{Murr}) that the relevant
parameter is $\rho = a/r$, where $r$ is the infectiveness and $a$ is the
removal rate.  We expect to verify the same for $\beta$ and $\tau$, so
the relevant parameter should be their product. This was numerically
checked for different values of $\beta$ and $\tau$ confirming the collapse
of all data. Consequently the critical promiscuity depends on the product
of $\beta$ and $\tau$ as can be observed in Fig~\ref{pc}, where
$\hat p^2_c$ as a function of $(\beta \tau)^{-1}$ is plotted along
with linear fittings for the three values of $\rho_{sng}$ studied.
From that we have obtained the expression
$\hat p^2_c \beta \tau = 0.36, 0.48, 0.65$ (respectively for
$\rho_{sng}=0, 0.5, 1$)
connecting the $\hat p$, $\beta$, and
$\tau$ parameters at the critical point.

As for practical purposes, we can test
the threshold condition, obtained from our model,
against HIV infection data. Assuming an infectivity
$\beta = 0.1$~\cite{Pet} and a life time $\tau = 5yr$, we can
roughly estimate $\hat p_c = 0.045$, this in turn means
($\langle p\rangle = (2/\pi)^{(1/2)}\hat p$ for a semi-Gaussian distribution)
an average critical promiscuity of $\langle p\rangle = 0.036$, i.e one ``excursion''
every 28 days in the population average. Thus, this estimate is
consistent with the outbreak of HIV infection among male
homosexuals in the late 70's in the US.

\begin{figure}[ht]
\includegraphics[width=8.cm, clip=true]{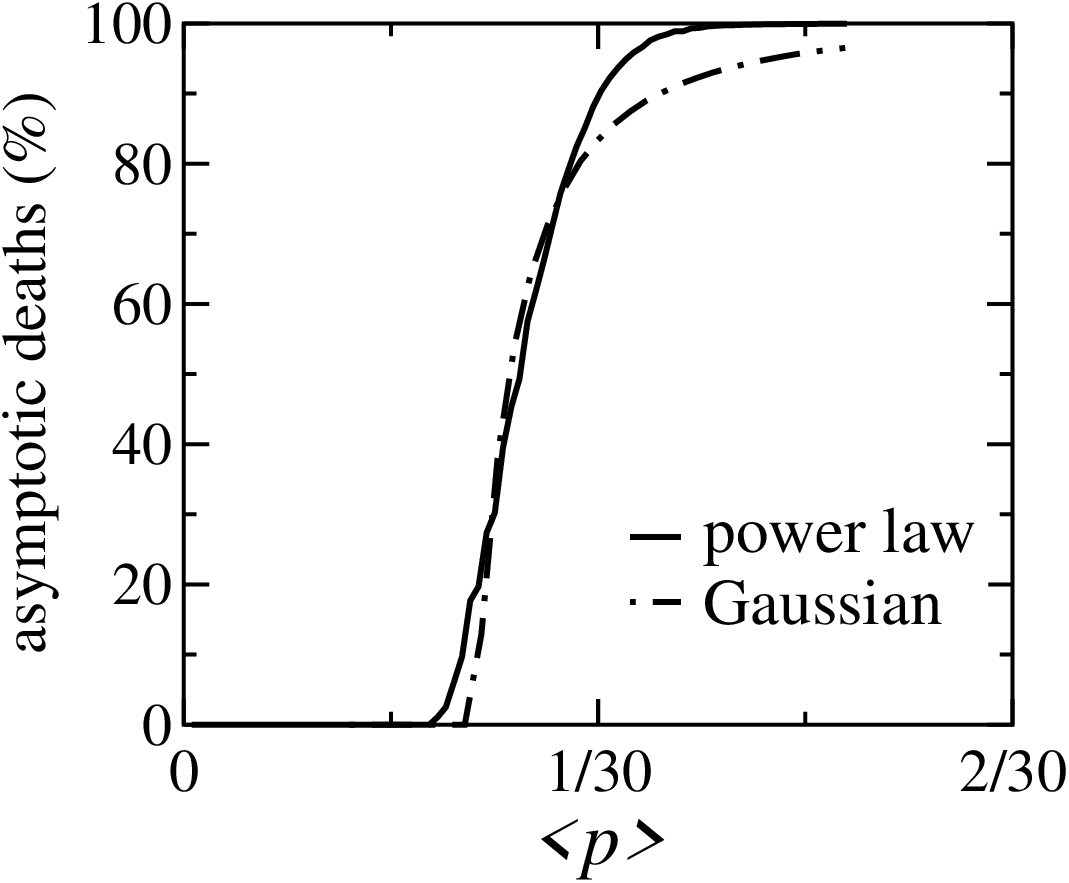}
\caption{
Asymptotic percentage of death as a function of the mean promiscuity
of the population $\langle p\rangle$ for a power-law distribution and a Gaussian
distribution of promiscuity. Parameters are $\tau=1yr$ and $\beta=1day^{-1}$.
Numerical results obtained with 100000 subjects,
100 realization and for $\rho_{sng}=0$ case (all subjects paired).}
\label{plaw}
\end{figure}

So far we have considered a semi-Gaussian distribution of promiscuity. It is
known today that the web of sexual contacts has a power low distribution as
shown in  ~\cite{Stan} for a sample of the swedish population, so the natural
question is: Are all the results presented here valid for an actual society or
are they mere academic speculation with no practical application? It has been
put forward for example that the scale-free structure of the sexual net make it
susceptible for an epidemic to occur without any threshold~\cite{SV}. In order
to verify this we have repeated some runs changing the semi-Gaussian
distribution by a power-law distribution:

   \[ n(p) = \left\{ \begin{array}{ll}
                      0              & \mbox{if  $p<p_0$} \\
                      2 p_0^2 p^{-3} & \mbox{otherwise}
                     \end{array}
             \right. \]

\noindent
and we have found again a threshold for the epidemic outbreak
For this distribution the parameter $p_0$ has to be bigger than  $0.1$ for an epidemic
to occur for $\beta=1$ and $\tau=1$.
The results can be observed in Fig.~\ref{plaw}, where we plot the asymptotic
number of death for the power law promiscuity distribution together with
the Gaussian distribution. In order to compare both distribution
we have normalized the promiscuity axis to the mean population
promiscuity $\langle p\rangle$. In this terms the onset of the epidemic are similar with
a shift to smaller values in the power law case.
Therefore the critical behavior is independent of the type of distribution.
This can be explained because of the distribution cut-off;
in the real world the net of contacts have no nodes (subjects) with infinity connectivity.
In our specific model, the maximum possible promiscuity
is 1 which means that the person goes out to meet someone everyday.
Even in this extreme case he has a maximum for the number of possible encounters
(less than 365 a year because he has to be accepted too).
We think that, at the same time that our hypothesis it reasonably for a real
population it gives an explanation for the apparent controversy
between our results and the theoretical arguments against the existence of
threshold~\cite{SV}.

\section{Conclusions}
We have presented a model for the spread of sexual transmitted
diseases, taking into account the social behavior. In spite of the
simplistic way of modeling the social interactions, the model
produce an outcome not previously reported in classical SIR
models for infectious diseases of any kind, sexually transmitted
or transmitted by other ways: a quantitative condition for the
epidemic to occur. This threshold, called by epidemiologist,
the basic reproductive number, was conjectured in the
classical literature in an empirical way~\cite{MA, ML}. Moreover,
it is not predicted, nor consistent with classical $SIR$ models.
On the other hand the present model gives a direct connection between
disease and social parameters and the epidemic threshold. Being
quantitative this connection could be of great utility to predict
epidemics. The relevant parameters are four:  two corresponding
to epidemiological aspects and related to the disease, the
probability of being infected in each sexual encounter $\beta$
and the average life time of the infectious individual $\tau$, and
two related to social aspects: the distribution of sexual
contacts of the population $p_i$, and the fraction of people that are
not in stable couples. The knowledge of the last three
would led us to determine the tolerated values for the
infectivity $\beta$ to slow down and, why not, to stop the HIV
infectious for example.

\begin{acknowledgments}
The authors thank Coordenação de Aperfeiçoamento de Pessoal de Nível Superior
(CAPES, Brasil) and Secretaría de Ciencia, Tecnología e Innovación Productiva
(SETCIP, Argentina) for support.
\end{acknowledgments}


\begin{thebibliography}{99}
\bibitem{Weid} W. Weidlich. Phys. Rep. {\bf 204}, 1 (1991).

\bibitem{MN} C. Moore and M. E. J. Newman, Phys. Rev. E {\bf 61}
(1999) 5678.

\bibitem{KA} M. Kuperman and G. Abramson, Phys. Rev. Lett. {\bf 86}, 2909 (2001).

\bibitem{SV} R. Pastor-Satorras and A. Vespignani, Phys. Rev. E {\bf
63}, 066117 (2001).

\bibitem{ZC} N. Zekri and J. P. Clerc, Phys. Rev. E {\bf 64},
056115 (2001).

\bibitem{SV2} R. Pastor-Satorras and A. Vespignani, Phys. Rev. Lett. {\bf
86}, 3200 (2001).

\bibitem{Newm} M. E. J. Newman, Phys. Rev. E {\bf 66}, 016128 (2002).

\bibitem{Murr} J. D. Murray, Mathematical Biology (Springer,
Berlin, 1993).

\bibitem{Pet} T. A. Peterman {\it et al.}, J. Am. med. Ass. {\bf 259}, 55 (1988).

\bibitem{MA} R. M. May and R. M. Anderson, Phil. Trans. R. Soc. Lond. {\bf B 321}, 565 (1988).

\bibitem{ML} R. M. May and A. L. Lloyd, Phys. Rev. E {\bf 64}, 066112
  (2001).

\bibitem{Stan} F. Liljeros, C. R. Edling, L. A. Nunes Amaral, H. E. Stanley,
Y. \AA berg, Nature {\bf 411}, 907 (2001).
\end{thebibliography}
\end{document}